\documentclass{revtex4}
\usepackage{amssymb}
\usepackage{amsfonts}
\usepackage{amsmath}

\input{tcilatex}

\begin{document}

\preprint{}
\title{Auxiliary quantization constraints on the von Roos ordering-ambiguity
at zero binding energies; azimuthally symmetrized cylindrical coordinates}
\author{Omar Mustafa}
\email{omar.mustafa@emu.edu.tr}
\affiliation{Department of Physics, Eastern Mediterranean University, G. Magusa, north
Cyprus, Mersin 10 - Turkey, Tel.: +90 392 6301078; fax: +90 3692 365 1604.}
\keywords{Position-dependent-mass, ordering-ambiguity parametric constraints.%
}

\begin{abstract}
\textbf{Abstract:} Using azimuthally symmetrized cylindrical coordinates, we
report the consequences of zero-energy quantal states on the von Roos
Hamiltonian. A position-dependent mass $M\left( \rho ,\varphi ,z\right)
=bz^{j}\rho ^{2\upsilon +1}/2$ is used. We show that the zero-energy setting
not only offers an additional degree of freedom towards feasible
separability for the von Roos Hamiltonian, but also manifestly yields
auxiliary quantized ambiguity parametric constraints.

\medskip
\end{abstract}

\pacs{PACS codes: 03.65.Ge, 03.65.Ca}
\maketitle

\section{Introduction}

Quantum mechanical potentials with zero binding energy (i.e., $E=0$) are
quantal states that represent (on their own mathematical solvability side)
quasi-exactly solvable quantum mechanical systems \cite{1}. Whilst, they
form the border line between the continuum and bound states of the energy
spectrum, they also offer exact solutions of Schr\"{o}dinger equation that
enlighten quantum-classical correspondence \cite{1,2,3,4,5,6,7}.. For
example, Daboul and Nieto \cite{2,3,4} have carried out systematic studies
on the power-law potentials and found several exceptional normalizable wave
functions (i.e., states that do not lie in the continuum spectral region)
for $E=0$, where the corresponding states are either bound or unbound.
Makowski and G\'{o}rska \cite{5}, have shown that the classical trajectories
of a particle precisely match with the localized quantum $E=0$ states.
Mazharimousavi \cite{7}, moreover, have reported the effects of
non-Hermitian $\mathcal{PT}$-symmetric settings on the localization of the $%
E=0$ states through his study of non-Hermitian quantum-classical
correspondence. On their applicability side, however, the $E=0$ states are
realized applicable in the cold-atom collisions, in the construction of some
vortex lattices, in the description of some modes in the Aharonov-Bohm
solenoids, and in quantum cosmology (cf, e.g., \cite{5,6,7} and the related
references cited therein).

On the other hand, position-dependent mass (PDM), $M\left( \vec{r}\right)
=m_{\circ }m\left( \vec{r}\right) $, quantum particles are described by the
von Roos Hamiltonian \cite{8} (with $m_{\circ }=\hbar =1$ units) 
\begin{equation}
H=-\frac{1}{4}\left[ m\left( \vec{r}\right) ^{\gamma }\vec{\nabla}m\left( 
\vec{r}\right) ^{\beta }\mathbf{\cdot }\vec{\nabla}m\left( \vec{r}\right)
^{\alpha }+m\left( \vec{r}\right) ^{\alpha }\;\vec{\nabla}m\left( \vec{r}%
\right) ^{\beta }\mathbf{\cdot }\vec{\nabla}m\left( \vec{r}\right) ^{\gamma }%
\right] +V\left( \vec{r}\right) ,
\end{equation}%
where, $\alpha $, $\beta $, and $\gamma $ are called the von Roos ordering
ambiguity parameters satisfying the von Roos constraint $\alpha +\beta
+\gamma =-1$ \cite%
{8,9,10,11,12,13,14,15,16,17,18,19,20,21,22,23,24,25,26,27,28,29,30,31,32,33,34,35,36,37,38,39,40,41,42,43}%
. The ordering ambiguity conflict is obviously manifested by the
non-uniqueness representation of the kinetic energy operator (cf., e.g., 
\cite{19}, \cite{33,34,35,36,37,38,39}). Nevertheless, in the search for
some physically acceptable parametric settings, it is found that the
continuity conditions at the abrupt heterojunction between two crystals
enforce the condition that $\alpha =\gamma $ (cf., e.g., Mustafa and
Mazharimousavi \cite{19} and Koc et al. \cite{37}). While the parametric
proposals of Ben Daniel and Duke ($\alpha =\gamma =0,$ $\beta =-1$), Zhu and
Kroemer ($\alpha =\gamma =-1/2,$ $\beta =0$), and Mustafa and Mazharimousavi
($\alpha =\gamma =-1/4,$ $\beta =-1/2$) \cite{19} satisfy this condition,
the Gora's and Williams' ($\beta =\gamma =0,$ $\alpha =-1$), and Li's and
Kuhn's ($\beta =\gamma =-1/2,$ $\alpha =0$) fail to do so. However, even
with this ordering ambiguity conflict arising in the process, L\.{e}%
vy-Leblond \cite{40} has advocated the correctness and conceptual
consistency of the use of position-dependent mass approximation approach.

In a recent work, Mustafa \cite{41} has considered the von Roos Hamiltonian
(1) using cylindrical coordinates and suggested a position-dependent mass
that is only radial-dependent (i.e., $m\left( \vec{r}\right) =m_{\circ
}M\left( \rho ,\varphi ,z\right) =M\left( \rho \right) =1/\rho ^{2}$) in
azimuthally symmetrized settings. Later on, Mustafa \cite{42} has offered a
parallel azimuthally symmetrized though a rather more general (but still
only radially-dependent) power-law-type position-dependent mass (i.e., $%
M\left( \rho ,\varphi ,z\right) =M\left( \rho \right) =b\rho ^{2\upsilon
+1}/2$). Moreover, spectral signatures of different $z$-dependent
interaction potential settings on the radial Coulombic and radial harmonic
oscillator interaction potentials' spectra were reported.

In the current methodical proposal, our motivation is inspired by a purely
theoretical mathematical curiosity as to what consequences would emerge at
zero-energy (i.e.,$E=0$) quantal states for the von Roos Hamiltonian (1).
Again, nevertheless, we use cylindrical coordinates in an azimuthally
symmetrized setting. We propose a position-dependent mass of yet a more
general and mixed dimensional-dependent form $M\left( \rho ,\varphi
,z\right) =bz^{j}\rho ^{2\upsilon +1}/2;$ $b,\,j,\,\upsilon \in 
%TCIMACRO{\U{211d} }%
%BeginExpansion
\mathbb{R}
%EndExpansion
$. We show that such $E=0$\ setting not only offers an additional degree of
freedom towards feasible separability of Hamiltonian (1), but also
manifestly yields quantized ordering ambiguity parametric constraints. To
the best of our knowledge, such position-dependent mass settings have not
been considered elsewhere.

This work is organized as follows. In section II, we recollect the most
relevant and necessary equations of \cite{42} (strictly speaking, equations
(2), (4), (5), (6), and (9) of \cite{42} summarized in (2)-(7) below, with $%
E=0$, of course). In so doing, we make the current work almost
self-contained. In the same section, we report on the separability of (1) as
a consequence of $E=0$ and provide the corresponding components in the
1D-Schr\"{o}dinger equation format. We show, in section III, that this
choice would introduce quantization recipes on the ordering ambiguity
parameters. Therein, we give illustrative examples of different interaction
potentials that although they may look complicated with mixed coordinates'
dependence, their exact solutions are simple and straightforward. The
corresponding wave functions are classified as textbook normalizable wave
functions. Our concluding remarks are given in section IV.

\section{Cylindrical coordinates at zero-energies and power-law PDM}

Following closely our recent works \cite{42} on cylindrical coordinates of
the PDM-Hamiltonian (1), we again consider the position-dependent mass and
the interaction potential to take the forms%
\begin{equation*}
m\left( \vec{r}\right) \equiv M\left( \rho ,\varphi ,z\right) =g\left( \rho
\right) f\left( \varphi \right) k\left( z\right) \text{ \ and }V\left( \vec{r%
}\right) \equiv V\left( \rho ,\varphi ,z\right) ,
\end{equation*}%
respectively. We have reported (see Mustafa \cite{42} for more details on
this issue) that the corresponding PDM-Schr\"{o}dinger equation%
\begin{equation*}
\left[ H-E\right] \Psi \left( \rho ,\varphi ,z\right) =0
\end{equation*}%
with%
\begin{equation}
\Psi \left( \rho ,\varphi ,z\right) =R\left( \rho \right) \Phi \left(
\varphi \right) Z\left( z\right) ;\text{ }\rho \in \left( 0,\infty \right)
,\ \varphi \in \left( 0,2\pi \right) ,z\in \left( -\infty ,\infty \right) ,
\end{equation}%
\begin{equation}
Z\left( z\right) =\sqrt{k\left( z\right) }\tilde{Z}\left( z\right) \text{, }%
\Phi \left( \varphi \right) =\sqrt{f\left( \varphi \right) }\tilde{\Phi}%
\left( \varphi \right) \text{, }
\end{equation}%
and%
\begin{equation}
g\left( \rho \right) =\frac{b}{2}\rho ^{2\upsilon +1},\text{and }R\left(
\rho \right) =\rho ^{\upsilon }U\left( \rho \right) ,
\end{equation}%
would (with $E=0$ in (11) of \cite{42}) imply%
\begin{align}
0& =\left[ \frac{U^{\prime \prime }\left( \rho \right) }{U\left( \rho
\right) }+\frac{\left( 2\upsilon +1\right) ^{2}\left[ \zeta -\beta -1\right]
-2\upsilon \left( \upsilon +1\right) }{2\rho ^{2}}-\tilde{V}\left( \rho
\right) \right]   \notag \\
& +\left[ \frac{\tilde{Z}^{\prime \prime }\left( z\right) }{\tilde{Z}\left(
z\right) }+\frac{\left( 2\zeta -3\right) }{4}\left( \frac{k^{\prime }\left(
z\right) }{k\left( z\right) }\right) ^{2}-\frac{\beta }{2}\frac{k^{^{\prime
\prime }}\left( z\right) }{k\left( z\right) }-\tilde{V}\left( z\right) %
\right]   \notag \\
& +\frac{1}{\rho ^{2}}\left[ \frac{\tilde{\Phi}^{\prime \prime }\left(
\varphi \right) }{\tilde{\Phi}\left( \varphi \right) }+\frac{\left( 2\zeta
-3\right) }{4}\left( \frac{f^{\prime }\left( \varphi \right) }{f\left(
\varphi \right) }\right) ^{2}-\frac{\beta }{2}\frac{f^{^{\prime \prime
}}\left( \varphi \right) }{f\left( \varphi \right) }-\tilde{V}\left( \varphi
\right) \right] .
\end{align}%
Where%
\begin{equation}
\zeta =\alpha \left( \alpha -1\right) +\gamma \left( \gamma -1\right) -\beta
\left( \beta +1\right) ,
\end{equation}%
and%
\begin{equation}
2MV\left( \rho ,\varphi ,z\right) =2g\left( \rho \right) f\left( \varphi
\right) k\left( z\right) V\left( \rho ,\varphi ,z\right) =\tilde{V}\left(
\rho \right) +\tilde{V}\left( z\right) +\frac{1}{\rho ^{2}}\tilde{V}\left(
\varphi \right) .
\end{equation}%
Consequently, the zero-energy, assumption secures separability of the von
Roos Hamiltonian (1) for some non-zero $k\left( z\right) =z^{j}$ of $M\left(
\rho ,\varphi ,z\right) =g\left( \rho \right) f\left( \varphi \right)
k\left( z\right) $ and hence a more general position-dependent mass form is
manifested in the process. That is, our position-dependent mass takes the
form (with $f\left( \varphi \right) =1$, $g\left( \rho \right) =b\rho
^{2\upsilon +1}/2$ and $k\left( z\right) =z^{j}$) as%
\begin{equation*}
M\left( \rho ,\varphi ,z\right) =bz^{j}\rho ^{2\upsilon +1}/2;b,j,\upsilon
\in 
%TCIMACRO{\U{211d} }%
%BeginExpansion
\mathbb{R}
%EndExpansion
\end{equation*}%
At this point, one should notice that choosing any other value for $E$
(i.e., $E\neq 0$) \ would make (11) of \cite{42} non-separable under our
current methodical proposal settings.

Next, we again choose to remain within azimuthal symmetrization settings and
consider that $\tilde{V}\left( \varphi \right) =0$ and $f\left( \varphi
\right) =1$ in (5) to imply that%
\begin{equation}
\frac{\tilde{\Phi}^{\prime \prime }\left( \varphi \right) }{\tilde{\Phi}%
\left( \varphi \right) }=k_{\varphi }^{2}\text{ ; }k_{\varphi }^{2}=-m^{2}%
\text{ ; }\left\vert m\right\vert =0,1,2,\cdots ,
\end{equation}%
where $m$ is the magnetic quantum number. Moreover, substituting $k\left(
z\right) =z^{j}$ in (5) yields that%
\begin{equation}
\left[ -\partial _{z}^{2}+\tilde{V}\left( z\right) +\frac{F\left( \alpha
,\beta ,\gamma ,j\right) }{z^{2}}\right] \,\tilde{Z}\left( z\right)
=k_{z}^{2}\,\tilde{Z}\left( z\right) ,
\end{equation}%
and%
\begin{equation}
\left[ -\partial _{\rho }^{2}+\frac{\tilde{\ell}_{\upsilon }^{2}-1/4}{\rho
^{2}}+\tilde{V}\left( \rho \right) \right] U\left( \rho \right)
=-k_{z}^{2}U\left( \rho \right) ,
\end{equation}%
where an irrational magnetic quantum number $\tilde{\ell}_{\upsilon }$ is
introduced in the process as 
\begin{equation}
\left\vert \tilde{\ell}_{\upsilon }\right\vert =\sqrt{\upsilon \left(
\upsilon +1\right) +m^{2}+\frac{1}{4}-\frac{\left( 2\upsilon +1\right) ^{2}%
\left[ \zeta -\beta -1\right] }{2}},
\end{equation}%
and%
\begin{equation}
F\left( \alpha ,\beta ,\gamma ,j\right) =-j\left[ j\left( \frac{2\zeta -3}{4}%
\right) -\left( j-1\right) \frac{\beta }{2}\right] \in 
%TCIMACRO{\U{211d} }%
%BeginExpansion
\mathbb{R}
%EndExpansion
.
\end{equation}%
Hereby, $F\left( \alpha ,\beta ,\gamma ,j\right) /z^{2}$ plays the role of a
manifestly repulsive and/or attractive force field. It is then convenient to
use the assumption that%
\begin{equation}
F\left( \alpha ,\beta ,\gamma ,j\right) =\mathcal{L}^{2}-1/4\Longrightarrow 
\mathcal{L}=\pm \sqrt{F\left( \alpha ,\beta ,\gamma ,j\right) +1/4}\in 
%TCIMACRO{\U{211d} }%
%BeginExpansion
\mathbb{R}
%EndExpansion
\end{equation}%
so that the condition%
\begin{equation*}
F\left( \alpha ,\beta ,\gamma ,j\right) +1/4\geq 0
\end{equation*}%
serves as an auxiliary constraint on the ambiguity parameters. Moreover, our
interaction potential takes the general form%
\begin{equation}
V\left( \rho ,\varphi ,z\right) =\frac{1}{bz^{j}\rho ^{2\upsilon +1}}\left[ 
\tilde{V}\left( \rho \right) +\tilde{V}\left( z\right) \right]
\,;b,j,\upsilon \in 
%TCIMACRO{\U{211d} }%
%BeginExpansion
\mathbb{R}
%EndExpansion
.
\end{equation}

In the following section, we use simple illustrative examples so that the
message of the current methodical proposal is made clear.

\section{$E=0$ and ambiguity parameters' quantization correspondence}

In this section, we show that when $E=0$, the ordering ambiguity parametric
constraints indulge quantization recipes. We use the exactly solvable
harmonic oscillator and the Coulomb potentials for constructive
illustrations.

A priori, let us provide exact solutions for the $z$-dependent equation in
(9) by recollecting the exact solutions for the harmonic oscillator, $\tilde{%
V}\left( z\right) =\tilde{a}^{2}z^{2}/4$, and for the Coulombic, $\tilde{V}%
\left( z\right) =-2\tilde{B}/z$, potentials. In so doing, we shall introduce
an impenetrable infinite wall for all $z<0$ and hence work in the upper-half
of the cylindrical coordinate system at hand. Mathematically speaking, we
propose 
\begin{equation}
\tilde{V}\left( z\right) :=\left\{ 
\begin{tabular}{l}
$\infty $ for $z<0$ \\ 
$\tilde{V}\left( z\right) $ for $z\geq 0$%
\end{tabular}%
\ \right. .
\end{equation}%
Under such settings and in a straightforward manner, one obtains%
\begin{equation}
k_{z}^{2}=-\sqrt{\tilde{a}^{2}}\left[ 2n_{z}+\left\vert \mathcal{L}%
\right\vert +1\right] 
\end{equation}%
for the harmonic oscillator, and%
\begin{equation}
k_{z}=\pm \frac{\tilde{B}}{\left( n_{z}+\left\vert \mathcal{L}\right\vert
+1\right) }
\end{equation}%
for the Coulombic interaction, where $\mathcal{L}$ is defined in (13). This
would immediately suggest that the corresponding wave functions are also
well-known exact solutions. They are, no doubt, the standard textbook
normalized wave functions of either the harmonic oscillator or the Coulomb
models.

In what follows, the constituents $\tilde{V}\left( \rho \right) $ and $%
\tilde{V}\left( z\right) $ of the interaction potential $V\left( \rho
,\varphi ,z\right) $ in (14) shall be chosen to be simple and exactly
solvable (Coulombic and/or harmonic oscillator) in their corresponding
radial (10) and/or $z$-coordinate (9) 1D Schr\"{o}dinger-like equations.
Therefore, the general forms of the corresponding wave functions are exact
and textbook normalized wave functions (given by (2) along with (3), and
(4)). Four complementary illustrative cases are in order.

\begin{description}
\item[Case 1] $\upsilon =1/2$, $\tilde{V}\left( \rho \right) =a^{2}\rho
^{2}/4,$and $\tilde{V}\left( z\right) =\tilde{a}^{2}z^{2}/4$.
\end{description}

The substitutions of $\upsilon =1/2$, $\tilde{V}\left( \rho \right)
=a^{2}\rho ^{2}/4,$ and $\tilde{V}\left( z\right) =\tilde{a}^{2}z^{2}/4$ in
(4) and (14) would suggest a position-dependent mass%
\begin{equation*}
M\left( \rho ,\varphi ,z\right) =bz^{j}\rho ^{2}/2
\end{equation*}%
moving under the influence of an interaction potential%
\begin{equation}
V\left( \rho ,\varphi ,z\right) =\frac{a^{2}}{4bz^{j}}+\frac{\tilde{a}^{2}}{%
4b\rho ^{2}z^{j-2}}.
\end{equation}%
Hence, the radial equation (10) yields%
\begin{equation}
k_{z}^{2}=-\sqrt{a^{2}}\left[ 2n_{\rho }+1+\sqrt{\left( m^{2}+3\right)
-2\left( \zeta -\beta \right) }\right] ,
\end{equation}%
and the $z$-dependent equation (9) implies that 
\begin{equation}
k_{z}^{2}=\sqrt{\tilde{a}^{2}}\left[ 2n_{z}+1+\sqrt{F\left( \alpha ,\beta
,\gamma ,j\right) +1/4}\right] ,
\end{equation}%
where $F\left( \alpha ,\beta ,\gamma ,j\right) $ is given in (12). If we
implement an over simplified assumption that $\sqrt{\tilde{a}^{2}}=-\sqrt{%
a^{2}}$, one would then use (19) and (20) to obtain%
\begin{equation}
-\frac{1}{4}+\left[ 2\left( n_{\rho }-n_{z}\right) +\sqrt{\left(
m^{2}+3\right) -2\left( \zeta -\beta \right) }\right] ^{2}=-j\left[ j\left( 
\frac{2\zeta -3}{4}\right) -\left( j-1\right) \frac{\beta }{2}\right] .
\end{equation}%
This is an auxiliary quantization constraint on the ordering ambiguity
parameters.

\begin{description}
\item[Case 2] $\upsilon =-1$, $\tilde{V}\left( \rho \right) =-2\,\tilde{A}%
\,/\rho $, and $\tilde{V}\left( z\right) =\tilde{a}^{2}z^{2}/4$
\end{description}

Such proposals in (4) and (14) would imply a position-dependent mass%
\begin{equation*}
M\left( \rho ,\varphi ,z\right) =bz^{j}\rho ^{-1}/2
\end{equation*}%
moving in an interaction potential of the form%
\begin{equation}
V\left( \rho ,\varphi ,z\right) =-\frac{2\,\tilde{A}}{bz^{j}}+\frac{\tilde{a}%
^{2}\rho }{4bz^{j-2}}.
\end{equation}%
In this case, the radial equation (10) yields 
\begin{equation}
k_{z}=\pm \frac{\tilde{A}}{\left( n_{\rho }+1+\sqrt{\left( m^{2}+3/4\right)
-\left( \zeta -\beta \right) /2}\right) },
\end{equation}%
and the $z$-dependent equation (9) gives%
\begin{equation}
k_{z}^{2}=\sqrt{\tilde{a}^{2}}\left[ 2n_{z}+1+\sqrt{F\left( \alpha ,\beta
,\gamma ,j\right) +1/4}\right] .
\end{equation}%
Therefore, 
\begin{equation}
-\frac{1}{4}+\left[ \frac{\tilde{A}^{2}/\left\vert \tilde{a}\right\vert }{%
\left( n_{\rho }+1+\sqrt{\left( m^{2}+\frac{3}{4}\right) -\frac{\left( \zeta
-\beta \right) }{2}}\right) ^{2}}-2n_{z}-1\right] ^{2}=-j\left[ j\left( 
\frac{2\zeta -3}{4}\right) -\left( j-1\right) \frac{\beta }{2}\right] .
\end{equation}%
Which is now the auxiliary quantization constraint on the ambiguity
parameters.

\begin{description}
\item[Case 3] $\upsilon =1/2$, $\tilde{V}\left( \rho \right) =a^{2}\rho
^{2}/4$, and $\tilde{V}\left( z\right) =-2\tilde{B}/z$
\end{description}

Such model suggests a position-dependent mass%
\begin{equation*}
M\left( \rho ,\varphi ,z\right) =bz^{j}\rho ^{2}/2
\end{equation*}%
moving in an interaction potential of the form%
\begin{equation}
V\left( \rho ,\varphi ,z\right) =\frac{a^{2}}{4bz^{j}}-\frac{2\tilde{B}}{%
b\rho ^{2}z^{j+1}}.
\end{equation}%
Therefore,%
\begin{equation}
k_{z}=\pm \left\vert a\right\vert \left[ 2n_{\rho }+1+\sqrt{\left(
m^{2}+3\right) -2\left( \zeta -\beta \right) }\right] ^{1/2},
\end{equation}%
\begin{equation}
k_{z}=\pm \frac{\tilde{B}}{\left( n_{z}+1+\sqrt{F\left( \alpha ,\beta
,\gamma ,j\right) +1/4}\right) },
\end{equation}%
and the auxiliary quantization constraint on the ambiguity parameters reads%
\begin{equation}
-\frac{1}{4}+\left[ \frac{\tilde{B}/\left\vert a\right\vert }{\left(
2n_{\rho }+1+\sqrt{\left( m^{2}+\frac{3}{4}\right) -\frac{\left( \zeta
-\beta \right) }{2}}\right) ^{\frac{1}{2}}}-n_{z}-1\right] ^{2}=-j\left[
j\left( \frac{2\zeta -3}{4}\right) -\left( j-1\right) \frac{\beta }{2}\right]
.
\end{equation}

\begin{description}
\item[Case 4] $\upsilon =-1$, $\tilde{V}\left( \rho \right) =-2\,\tilde{A}%
\,/\rho $, and $\tilde{V}\left( z\right) =-2\tilde{B}/z$
\end{description}

This yields a position-dependent mass of the form%
\begin{equation*}
M\left( \rho ,\varphi ,z\right) =bz^{j}\rho ^{-1}/2
\end{equation*}%
and an interaction potential%
\begin{equation}
V\left( \rho ,\varphi ,z\right) =-\frac{2\,\tilde{A}}{bz^{j}}-\frac{2\tilde{B%
}\rho }{bz^{j+1}}.
\end{equation}%
Hence, 
\begin{equation}
k_{z}=\pm \frac{\tilde{A}}{\left( n_{\rho }+1+\sqrt{\left( m^{2}+3/4\right)
-\left( \zeta -\beta \right) /2}\right) },
\end{equation}%
\begin{equation}
k_{z}=\pm \frac{\tilde{B}}{\left( n_{z}+1+\sqrt{F\left( \alpha ,\beta
,\gamma ,j\right) +1/4}\right) },
\end{equation}%
and the auxiliary quantization constraint on the ambiguity parameters is 
\begin{equation}
-\frac{1}{4}+\left[ \frac{\tilde{B}}{\tilde{A}}\left( n_{\rho }+1+\sqrt{%
\left( m^{2}+\frac{3}{4}\right) -\frac{\left( \zeta -\beta \right) }{2}}%
\right) -n_{z}-1\right] ^{2}=-j\left[ j\left( \frac{2\zeta -3}{4}\right)
-\left( j-1\right) \frac{\beta }{2}\right] .
\end{equation}

In the four cases discussed above, we observe auxiliary quantization
constraints on the ordering ambiguity parameters (documented in the
appearance of $n_{\rho },$ $n_{z}$, and $m$ quantum numbers in (21), (25),
(29), and (33)). That is, for each set of values of $n_{\rho },$ $n_{z}$,
and $m$ there is a corresponding quantized ordering ambiguity parametric
constraint. For example, in Case 1 and for $j=0$ in (21), one obtains $%
F\left( \alpha ,\beta ,\gamma ,j\right) =0$ and%
\begin{equation}
\zeta -\beta =\frac{1}{2}\left[ m^{2}+3-\left( 2n_{z}-2n_{\rho }+\frac{1}{2}%
\right) ^{2}\right] ,
\end{equation}%
where $\zeta $ is given in (6). It is obvious here that for every set of
values of $n_{\rho },$ $n_{z}$, and $m$ the profile of the ordering
ambiguity parameters would change and consequently the profile of the
kinetic energy operator in (1) would change. Hence the effective potential
of (1) would change its profile too. Moreover, one observes that for $%
n_{\rho }=n_{z}=\left\vert m\right\vert =0$, this quantized constraint (34)
is only satisfied by the set of $\alpha =\gamma =-1/4$ and $\beta =-1/2$ of
Mustafa and Mazharimousavi \cite{19}. Such a result does not make this
parametric set as a universally acceptable one, of course. Changing the
quantum numbers $n_{\rho },$ $n_{z}$, and $m$ would change the profile of
the acceptable parametric sets. The same trend may very well be observed in
Cases 2, 3, and 4. The ordering ambiguity constraints' quantization
correspondence, as an obvious manifestation of $E=0$ quantal states, is
therefore clear.

\section{Concluding remarks}

Under azimuthally symmetric settings, we have recollected the most relevant
and vital relations (equations (2)-(7) above) that have been readily
reported by Mustafa \cite{42} for cylindrical coordinates separability of
the von Roos Hamiltonian (1). Therein \cite{42}, the position-dependent mass 
$M\left( \rho ,\varphi ,z\right) =g\left( \rho \right) =b\rho ^{2\upsilon
+1}/2;\upsilon ,b\in 
%TCIMACRO{\U{211d} }%
%BeginExpansion
\mathbb{R}
%EndExpansion
$ is introduced as a generalization of $M\left( \rho ,\varphi ,z\right)
=g\left( \rho \right) =1/\rho ^{2}$ of \cite{41}. Spectral signatures of
different $z$-dependent interaction potential settings on the radial
Coulombic and radial harmonic oscillator interaction potentials' spectra
were reported.

In the current methodical proposal, however, we have discussed the
consequences of choosing zero-energy states ( i.e., states with $E=0$) for
our position-dependent mass Hamiltonian (1) under azimuthally symmetrized
cylindrical coordinates settings. Moreover, we have used a more general
position-dependent mass function, $M\left( \rho ,\varphi ,z\right)
=bz^{j}\rho ^{2\upsilon +1}/2;$ $b,\,j,\,\upsilon \in 
%TCIMACRO{\U{211d} }%
%BeginExpansion
\mathbb{R}
%EndExpansion
$. We have shown that the choice of $E=0$\ setting provides not only an
additional degree of freedom towards the feasible separability of
Hamiltonian (1), but also manifestly yields quantized ordering ambiguity
parametric constraints (documented in (21), (25), (29), (33), and (34)). We
have also shown that even with the simplistic choices of the constituent
interaction potentials $\tilde{V}\left( \rho \right) $ and $\tilde{V}\left(
z\right) $, the overall general interaction potentials, $V\left( \rho
,\varphi ,z\right) $, turned out to be complicated in the sense of indulging
mixed coordinates dependence (documented in (18), (22), (26), and (30)).
Yet, their exact solutions are simple and straightforward. They are the
exact and textbook normalizable wave functions of either the harmonic
oscillator or the Coulomb models in (9) and (10). Consequently, the overall
general forms of the wave functions are exact, textbook normalizable, and
given through (2), (3), and (4).

\end{document}